\documentclass[journal,12pt,onecolumn,draftclsnofoot,]{IEEEtran}

\usepackage{cite}  
\usepackage{bm} 
\usepackage[pdftex]{graphicx}
\usepackage{stfloats} 
\usepackage{amsmath,amsfonts,amssymb,amscd,empheq}
\usepackage{hyperref}

\usepackage{multirow}
\usepackage{bbm}
\usepackage{bbold}

\usepackage{amsthm}
\usepackage{upgreek}

\usepackage{setspace}
\usepackage{pdfsync}
\usepackage{epstopdf}
\usepackage{lipsum,multicol}
\usepackage[numbered,autolinebreaks,useliterate]{mcode}
\usepackage{mathtools,bbm}
\usepackage{mathtools, dsfont}

\usepackage{caption}
\usepackage{subcaption}
\usepackage{scalefnt}

\usepackage{lipsum}
\usepackage{calligra}
\DeclareMathAlphabet{\mathcalligra}{T1}{calligra}{m}{n}

\DeclareMathOperator{\argmin}{argmin}

\newcommand{\vertiii}[1]{{\left\vert\kern-0.25ex\left\vert\kern-0.25ex\left\vert #1 
    \right\vert\kern-0.25ex\right\vert\kern-0.25ex\right\vert}}

\usepackage[noend]{algpseudocode}
\usepackage{algorithm,algorithmicx}

\algrenewcommand\algorithmicrequire{\textbf{Initialization:}}
\algrenewcommand\algorithmicensure{\textbf{For each $i$, execute for $k\geq 0$:}}

\algblockdefx[Loop]{Loop}{EndLoop}[1]{\textbf{for $k\geq 0$:} #1}

\makeatletter
\algrenewcommand\ALG@beginalgorithmic{\small}
\makeatother

\makeatletter
\newcommand\fs@betterruled{%
  \def\@fs@cfont{\bfseries}\let\@fs@capt\floatc@ruled
  \def\@fs@pre{\vspace*{5pt}\hrule height.8pt depth0pt \kern2pt}%
  \def\@fs@post{\kern2pt\hrule\relax}%
  \def\@fs@mid{\kern2pt\hrule\kern2pt}%
  \let\@fs@iftopcapt\iftrue}
\floatstyle{betterruled}
\restylefloat{algorithm}
\makeatother

\newcommand*{\vsepfbox}[1]{%
  \begingroup
    \sbox0{\fbox{#1}}%
    \setlength{\fboxrule}{0pt}%
    \mbox{\kern-\fboxsep\fbox{\unhbox0}\kern-\fboxsep}%
  \endgroup
}

\usepackage{cases}
\usepackage{balance}
\usepackage{float}
\usepackage{tabu}

\allowdisplaybreaks
\usepackage{enumitem}
\setlist[itemize]{align=parleft,left=0pt..1em}
\begin{document}
\title{{Data-Driven Secondary Control of Distributed Energy Resources}}

\author{Madi Zholbaryssov and Alejandro D. Dom\'inguez-Garc\'ia,~\IEEEmembership{Senior Member,~IEEE}
\thanks{The authors are with the Department of Electrical and Computer Engineering of the University of Illinois at Urbana-Champaign, Urbana, IL 61801,
USA. Email: {\tt\small \{zholbar1, aledan\}@ILLINOIS.EDU.}
}}

\maketitle
\begin{abstract}
In this paper, we present a data-driven secondary controller for regulating to some desired values several variables of interest in a power system, namely, electrical frequency, voltage magnitudes at critical buses, and active power flows through critical lines. The power generation system is based on distributed energy resources (DERs) interfaced with either grid-forming (GFM) or grid-following (GFL) inverters. The secondary controller is based on online feedback optimization leveraging the learned sensitivities of the changes in the system frequency, voltage magnitudes at critical buses, and active power flows through critical lines to the changes in inverter active and reactive power setpoints. To learn the sensitivities accurately from data, the feedback optimization has a built-in mechanism for keeping the secondary control inputs persistently exciting without degrading its performance. The feedback optimization also utilizes the learned power-voltage characteristics of photovoltaic (PV) arrays to compute DC-link voltage setpoints so as to allow the PV arrays to track the power setpoints. To learn the power-voltage characteristics, we separately execute a data-driven approach that fits a concave polynomial to the collected power-voltage measurements by solving a sum-of-squares (SoS) optimization. We showcase the secondary controller using the modified IEEE-$14$ bus test system, in which conventional energy sources are replaced with inverter-interfaced DERs.
\end{abstract}

\IEEEpeerreviewmaketitle

\section{Introduction}
With the increased installation of smart meters and other measurement devices in power distribution systems, there is a growing impetus for designing cost-effective and reliable control methods that harness the full potential of the data gathered by such measurement devices. In particular, data-driven approaches can help address the problem of integration of intermittent renewable energy sources (RESs) into power grids \cite{ZhYa16}.
To achieve the ultimate goal of replacing conventional fossil fuel-based energy sources with RESs, intermittency of their power output must be considered in the control system design, e.g., by endowing RESs with fast frequency and voltage regulation capabilities. Availability of data and recent advances in the area of data-driven decision-making (see, e.g, \cite{PeTe20}) offer us new opportunities and tools for dealing with uncertainty associated with the power produced by RESs.

Different data-driven techniques for power systems control and operation have recently been developed in the literature (see, e.g., \cite{ChDo14,HoDo16,GoTo19,XuDo19,KaHu19,XuDo20,MaKa20,ChZh20,YaSu20}). A number of works (see, e.g., \cite{GoTo19,MaKa20}) proposed data-driven controllers that solve some underlying optimization problem, using measurements to perform either system identification or controller synthesis directly without system identification. The authors of \cite{ChDo14} propose a method to compute linear
sensitivity distribution factors in near real-time, which was utilized in \cite{HoDo16} to perform a real-time security-constrained economic dispatch.
The authors of \cite{GoTo19} propose a data-driven approach for frequency control of power systems with variable inertia, which is based on designing a static feedback gain from the datasets generated by the linear quadratic regulator (LQR) for different modes, where each mode depends on the amount of inertia present in the system. The authors of \cite{XuDo19} propose a data-driven approach for regulating the active power exchange between a power distribution system and the bulk grid to which it is interconnected, which uses the learned sensitivities; loss of the persistent excitation in the DER power setpoints is prevented by adding a noise sampled from the Bernoulli distribution. The authors of \cite{XuDo20} propose a data-driven approach for voltage regulation in radial power distribution systems based on the estimated linear branch flow model; however, the approach requires the system to be fully observable, i.e., each bus must be equipped with a meter that provides voltage and power injection measurements. The authors of \cite{MaKa20} propose a data-driven $H_2$ optimization to design a non-droop-based controller for achieving the standard primary and secondary control objectives, which directly uses measurements for controller synthesis, thereby bypassing the need for system identification. Another body of works (see, e.g., \cite{ChZh20,YaSu20}) utilized emerging machine learning techniques, e.g., deep learning, deep reinforcement learning, to make control decisions based on deep neural networks that replace uncertain power system models; however, training deep neural networks requires large volumes of data.

In this paper, we propose a data-driven secondary controller for an inverter-based power system, the objective of which is to regulate to some desired values several variables of interest, namely, the electrical frequency across the system, voltage magnitudes at critical buses, and active power flows through critical lines. The design of the secondary controller is based on online feedback optimization that makes use of learned sensitivities of changes in the variables to be regulated to changes in the control inputs, comprising the active and reactive power setpoints of the inverters. These sensitivities are estimated recursively using real-time measurements. One of the main challenges in designing such controller lies in maintaining persistent excitation of the control inputs and preventing the loss of identifiability of the sensitivities due to feedback control. A common strategy to maintain the excitation is to add a random signal \cite{AW:08}. In this work, we argue that there is a better approach that relies on a certain mechanism built into the feedback optimization that adds a noise less detrimental to the control objectives only when necessary, and exploits the natural excitation of the output error due to, e.g., load fluctuations. 

An additional feature of the secondary controller is to enable PV arrays to provide regulation services. One of the challenges here is to compute the power setpoints that are trackable by the PV arrays given the fluctuations in their power-voltage characteristics due to the changes in solar irradiance. In this work, we attempt to integrate the learned power-voltage characteristics of the PV arrays into the proposed feedback optimization, and compute the DC-link voltage setpoints in addition to the power setpoints. In other words, the feedback optimization combines the standard secondary control objectives with the PV power tracking objectives.
Meanwhile, the power-voltage characteristics are modeled by a concave polynomial fit to the collected measurements by solving a SoS optimization \cite{ZhDh20}.

\section{Preliminaries}\label{sec:preliminaries}
In this section, we describe the adopted models of the power network and inverter-interfaced DERs, and formulate the secondary control problem under uncertainty in photovoltaic generation.
\subsection{Power Network}
We consider a collection of inverter-interfaced DERs and loads physically interconnected by an electrical network operating in balanced three-phase regime. Assume the network has $n$ buses indexed by the elements in the set $\mathcal{V}=\{1,2,\dots,n\}$. Assume the network has $m$ inverter-interfaced DERs indexed by the elements in the set $\mathcal{V}^{(g)}=\{1,2,\dots,m\}$. Let $\mathcal{V}_{pv}\subseteq\mathcal{V}^{(g)}$ denote the set of PV arrays interfaced with GFL inverters; in the remainder, we refer to these as solar inverters. The remaining DERs from the set $\mathcal{V}^{(I)}\coloneqq\mathcal{V}^{(g)}\setminus \mathcal{V}_{pv}$ are assumed to be interfaced with three-phase GFM inverters.

Let $\theta_i(t)$ and $V_i(t)$ respectively denote the phase and magnitude of the phasor associated with the voltage at bus $i$ at time $t$. Let $\omega_i(t)\coloneqq \frac{d\theta_i(t)}{dt}$ denote the local frequency at bus $i$ at time $t$. Let $V_i^*$ denote the nominal value of the magnitude of the phasor associated with the voltage at bus $i$, and let $\omega^*$ denote the system nominal frequency. 
Let $\omega(t)\coloneqq \frac{1}{m}\sum_{i\in\mathcal{V}^{(I)}}\omega_i(t)$ denote the average frequency, which is to be regulated by the secondary controller to the nominal value, $\omega^*$.
Let $\mathcal{B}$ denote the set of certain critical buses, at which the voltage magnitudes are to be regulated to their respective nominal values. Define $V(t)\coloneqq \big[\{V_i(t)\}_{i\in\mathcal{B}}\big]^\top$ and $V^*\coloneqq \big[\{V_i^*\}_{i\in\mathcal{B}}\big]^\top$. Let $\mathcal{T}$ denote the set of certain critical lines, e.g., tie lines, in which the active power flows are to be regulated to their respective nominal values. Let $(i,j)\in\mathcal{T}$ denote an electrical line connecting buses $i$ and $j$, $p_{ij}(t)$ denote the amount of active power that flows from bus $i$ to bus $j$ through line $(i,j)$, and $p_{ij}^*$ denote the corresponding nominal value. Define $p(t)\coloneqq \big[\{p_{ij}(t)\}_{(i,j)\in\mathcal{T}}\big]^\top$ and $p^*\coloneqq \big[\{p_{ij}^*\}_{(i,j)\in\mathcal{T}}\big]^\top$.

\subsection{Inverter Models}
In this work, three-phase GFM inverters are controlled using the following primary voltage and frequency droop-like control laws based on the dynamics of the so-called Andronov-Hopf oscillator \cite{LuJo19}:
\begin{subequations}\label{voc}
\begin{align}
\dot{E}_j(t) &= \frac{\xi_j}{\kappa_{j,v}^2}E_j(t)\big(2E_j^{*2} - 2E_j(t)^2\big) \nonumber\\ &\quad- \frac{\kappa_{j,v}\kappa_{j,i}}{3C_jE_j(t)}\big(Q_j(t)-Q_j^r(t)\big),\\
\dot{\delta_j}(t) &= \omega^* - \frac{\kappa_{j,v}\kappa_{j,i}}{3C_jE_j(t)^2}\big(P_j(t)-P_j^r(t)\big),\mbox{ }j\in \mathcal{V}^{(I)},
\end{align}
\end{subequations}
where $\delta_j(t)$ and $E_j(t)$ respectively denote the phase and magnitude of the phasor associated with the terminal voltage of the GFM inverter $j\in \mathcal{V}^{(I)}$ at time $t$, $E_j^{*}$ denotes the nominal inverter voltage, $\xi_j$, $\kappa_{j,i}$, $\kappa_{j,v}$, and $C_j$ are constants, $P_j(t)$ and $Q_j(t)$ denote the active and reactive power injected into the network by the inverter, $P_j^r(t)$ and $Q_j^r(t)$ denote the active and reactive power setpoints provided to the inverter by the secondary controller. 

GFL inverter $j\in\mathcal{V}_{pv}$ is capable of regulating its active and reactive power outputs denoted by $P_j(t)$ and $Q_j(t)$, respectively, to the corresponding active and reactive power setpoints denoted by $P_j^r(t)$ and $Q_j^r(t)$, provided that $P_j^r(t)$ does not exceed the active power capacity of the PV array. If $P_j^r(t)$ exceeds the active power capacity, then, the inverter is assumed to provide the maximum available active power.  

\subsection{PV Power Tracking Mechanism}\label{subsec:pv_tracking}

To extract a desired amount of power from a PV array $i\in\mathcal{V}_{pv}$, as specified by the secondary control setpoints, a certain voltage, denoted by $v_i$, must be applied across its terminals, as determined by its power-voltage characteristic. Hence, there is a need for a mechanism capable of converting the power setpoints computed by the secondary controller into the voltage setpoints tracked by the DC-side controller. On the other hand, the secondary controller must take into account the power-voltage characteristics of the PV arrays to ensure that the computed power setpoints are trackable by the solar inverters without exceeding the maximum active power capacity of the PV arrays.
A main challenge in designing such mechanism lies in dealing with the uncertainty of the power-voltage characteristics of the PV arrays. However, as shown in \cite{ZhDh20} and described next, these power-voltage characteristics can be accurately modeled by a degree-$d$ polynomial fitted to the collected power-voltage measurements.
Let $v_i(t)$ and $p_i(t)$ respectively denote the voltage applied across the terminals of the PV array $i\in\mathcal{V}_{pv}$ and the resulting active power output at time $t$. Then, we can write a polynomial for predicting the active power output of the PV array for a given terminal voltage $v_i$ as follows:
\begin{align}
c_i(v_i,\beta^{(i)}[k]) = \sum_{l=0}^{d}\beta_l^{(i)}[k]v_i^l.\label{poly_model}
\end{align}
The coefficients of the polynomial, $\beta^{(i)}[k] = [\beta^{(i)}_0[k],\beta^{(i)}_2[k],\dots,\beta^{(i)}_{d}[k]]^\top$, are estimated at time instant $t_k$ by fitting the model to the power-voltage measurements, $\{(v_i[l],p_i[l])\}_{l=k-w}^k$, collected over a time window denoted by $w$, where $v_i[l] \coloneqq v_i(t_l)$ and $p_i[l] \coloneqq p_i(t_l)$. The model fitting amounts to solving a linear regression problem with the additional constraints to ensure that the resulting polynomial is concave \cite{ZhDh20}.

\subsection{Control Objectives}

Let $\{t_k\}_{k\geq1}$ denote the sequence of time instants at which $P_j^r(t)$ and $Q_j^r(t)$, $j\in\mathcal{V}^{(g)}$, are adjusted. Let $P^r[k] \coloneqq \big[P_1^r[k],P_2^r[k],\dots,P_m^r[k]\big]^\top$, and $Q^r[k] \coloneqq \big[Q_1^r[k],Q_2^r[k],\dots,Q_m^r[k]\big]^\top$, where $P_j^r[k]\coloneqq P_j^r(t_k)$ and $Q_j^r[k]\coloneqq Q_j^r(t_k)$, and let $u[k] \coloneqq \big[P^r[k]^\top,Q^r[k]^\top\big]^\top$.
Then, our goal is to develop a secondary controller that regulates the output of the system, $y(t)\coloneqq \big[\omega(t),V(t)^\top,p(t)^\top\big]^\top$, to the nominal level, denoted by $y^*\coloneqq \big[\omega^*,V^{*\top},p^{*\top}\big]^\top$, by adjusting the secondary control inputs, $u[k]$, while taking into account the learned power-voltage characteristics in \eqref{poly_model} of the PV arrays.

\section{System Output Prediction Model}\label{sec:prediction_model}
In this section, we present the model for predicting the changes in the system output given the changes in the secondary control inputs. The model will later be used by the secondary controller (to be presented in Section~\ref{sec:sec_control}).

\subsection{Online Sensitivity Estimator}
Let $\{t_k^-\}_{k\geq1}$ denote the sequence of time instants such that $t_{k-1}<t_k^-<t_k$, $\forall k$, at which the output $y(t)$ is measured. Let $y[k]\coloneqq\big[\omega[k],V[k]^\top,p[k]^\top\big]^\top$ denote the measurement of the output $y(t)=\big[\omega(t),V(t)^\top,p(t)^\top\big]^\top$ at time $t = t_{k+1}^-$, where $\omega[k]\coloneqq\omega(t_{k+1}^-)$, $V[k]\coloneqq \big[\{V_i[k]\}_{i\in\mathcal{B}}\big]^\top$, with $V_j[k]\coloneqq V_j(t_{k+1}^-)$, and $p[k]\coloneqq \big[\{p_{ij}[k]\}_{(i,j)\in\mathcal{T}}\big]^\top$, with $p_{ij}[k]\coloneqq p_{ij}(t_{k+1}^-)$.
Define $\Delta P^r[k]\coloneqq P^r[k]-P^r[k-1]$, $\Delta Q^r[k]\coloneqq Q^r[k]-Q^r[k-1]$, and let $\Delta u[k] \coloneqq \big[\Delta P^r[k]^\top,\Delta Q^r[k]^\top\big]^\top$
denote the vector of the changes in the secondary control inputs producing the changes in the output denoted by $\Delta y[k]\coloneqq y[k]-y[k-1]$. Here, $y[k]$ can be expressed in terms of some nonlinear function $h(\cdot)$ as follows:
\begin{align}
y[k] = h(u[k],\chi[k]),
\end{align}
where $h(\cdot)$ depends on $u[k]$ and $\chi[k]\coloneqq \chi(t_k)$, with $\chi(t)$ denoting the exogenous disturbances representing electrical loads and unknown power network parameters. Then, by using the Taylor series expansion, and keeping only the linear term, we obtain that
\begin{align}
\Delta y[k] \approx S[k]\Delta u[k] + \epsilon[k],
\end{align}
where $S[k] \coloneqq \frac{\partial h(u[k-1],\chi[k-1])}{\partial u}$ is referred to as the sensitivity matrix, and $\epsilon[k] \coloneqq \frac{\partial h(u[k-1],\chi[k-1])}{\partial \chi}\Delta \chi[k]$. Consider the following linear prediction model:
\begin{align}
\widehat{\Delta y}[k] = \widehat{S}[k]\Delta u[k],
\end{align}
where $\widehat{\Delta y}[k]$ is the estimate of $\Delta y[k]$, and $\widehat{S}[k]$ denotes the estimate of the sensitivity matrix $S[k]$, which is estimated using the method of least squares as follows:
\begin{align}\label{ls_problem}
\widehat{S}[k+1] = \argmin\limits_{S} \sum_{l=1}^k\lambda^{k-l}\big\|\Delta y[l] - S\Delta u[l]\big\|_2^2,
\end{align}
where $\lambda \in (0,1)$ denotes the forgetting factor that allows the estimator to assign exponentially less weight to older measurements. The solution to the least-squares problem \eqref{ls_problem} cannot be uniquely identified at time instant $t_{k+1}$ unless $\{\Delta u[l]\}_{l=1}^k$ is persistently exciting \cite{AW:08}, namely, there exist an integer $\sigma<k$ and constants $\varrho_1,\varrho_2>0$ such that\footnote{We write $A>B$ ($A\geq B$) to denote that a matrix $A-B$ is positive definite (positive semidefinite).} 
\begin{align}\label{persistent_excitation}
    \varrho_1I>\sum_{l=k-\sigma}^{k} \Delta u[l]\Delta u[l]^\top > \varrho_2I;
\end{align}
then, in this case, the solution to \eqref{ls_problem} is given by
\begin{align}\label{ls_soln}
    \widehat{S}[k+1] = \Delta Y[k]^\top\Lambda[k] X[k]\big(X[k]^\top\Lambda[k] X[k]\big)^{-1},
\end{align}
where $X[k] = \big[\Delta u[1],\Delta u[2],\dots,\Delta u[k]\big]^\top$, $Y[k] = \big[\Delta y[1],\Delta y[2],\dots,\Delta y[k]\big]^\top$,
and $\Lambda[k]$ is a ($k\times k$)-dimensional diagonal matrix with $\Lambda_{ii}[k]=\lambda^{k-i}$. 

When computing the least-squares solution in \eqref{ls_soln}, we can avoid the need to perform matrix inversion by using the standard recursive least squares algorithm (see, e.g., \cite{AW:08}) given by
\begin{subequations}\label{rls}
\begin{align}
H[k] &= \frac{\lambda^{-1}}{1 + \lambda^{-1}\Delta u[k]^\top F[k]\Delta u[k]}F[k]\Delta u[k],\\
\widehat{S}[k+1] &= \widehat{S}[k] + H[k]\big(\Delta y[k] - \widehat{S}[k]\Delta u[k]\big),\\
F[k+1] &= \lambda^{-1}F[k] - \lambda^{-1}H[k]\Delta u[k]^\top F[k],
\end{align}
\end{subequations}
where $F[1] = \rho_1I$, with $\rho_1>0$. The initialization of $\widehat{S}[1]$ is discussed next.  

\subsection{Initialization of the Sensitivity Matrix Estimate}
Let $d_v\coloneqq |\mathcal{B}|$ and $d_l\coloneqq |\mathcal{T}|$ denote the lengths of $V[k]$ and $p[k]$, respectively. Let $S_f$, $S_v$, and $S_l$ denote the $(1\times 2m)$, $(d_v\times 2m)$, and  $(d_l\times 2m)$-dimensional submatrices of $\widehat{S}[1]$ comprised of the rows that correspond respectively to the frequency, voltages, and line flows being estimated; then, we have that $\widehat{S}[1] = [S_f^\top,S_v^\top,S_l^\top]^\top.$
Setting $\dot{E}_j(t) =0$ and $\dot{\delta}_j(t) =\omega(t)$ in \eqref{voc}, we obtain the voltage and frequency relations at steady state, which are given by \cite{LuJo19}:
\begin{subequations}\label{voc_eq}
\begin{align}
E_j[\tau] &= \frac{E_j^*}{\sqrt{2}}\left(1+\sqrt{1 - \frac{2\kappa_{j,v}^3\kappa_{j,i}}{3C_j\xi_jE_j^{*4}}\big(Q_j[\tau]-Q_j^r[\tau]\big)}\right)^{0.5},\\
\omega[\tau] &= \omega^* - \frac{\kappa_{j,v}\kappa_{j,i}}{3C_jE_j[\tau]^2}\big(P_j[\tau]-P_j^r[\tau]\big),
\end{align}
\end{subequations}
$j\in\mathcal{V}^{(I)}$, $\tau=1,2,\dots$, where $P_j[\tau]\coloneqq P_j(t_{\tau+1}^-)$, and $Q_j[\tau]\coloneqq Q_j(t_{\tau+1}^-)$.
By linearizing \eqref{voc_eq} around the nominal values, we obtain that
\begin{subequations}\label{voc_eq_lin}
\begin{align}
E_j[\tau] &= E_j^* - R_j^{(v)}(Q_j[\tau]-Q_j^r[\tau]),\\
\omega[\tau] &= \omega^* - R_j^{(f)}(P_j[\tau]-P_j^r[\tau]),
\end{align}
\end{subequations}
where $R_j^{(v)}\coloneqq {\kappa_{j,v}^3\kappa_{j,i}}/{(12C_j\xi_j E_j^{*3})}$ and $R_j^{(f)}\coloneqq {\kappa_{j,v}\kappa_{j,i}}/{(3C_jE_j^{*2})}$ are interpreted as voltage and frequency droops, respectively, determining steady-state voltage and frequency deviations from their nominal values resulting from changes in active and reactive power outputs.
Subtracting \eqref{voc_eq_lin} at $\tau=k-1$ from \eqref{voc_eq_lin} at $\tau=k$, we obtain that
\begin{subequations}
\begin{align}
D_{j,v}\Delta E_j[k]-\Delta Q_j^r[k]+\Delta Q_j[k]=0,\label{voc_eq_lin1}\\
D_{j,f}\Delta\omega[k]-\Delta P_j^r[k]+\Delta P_j[k]=0,\label{voc_eq_lin2}
\end{align}
\end{subequations}
where $D_{j,v} = 1/R_j^{(v)}$, $D_{j,f} = 1/R_j^{(f)}$, $\Delta E_j[k]\coloneqq E_j[k]- E_j[k-1]$, $\Delta\omega[k]\coloneqq\omega[k]-\omega[k-1]$, $\Delta P_j[k]\coloneqq P_j[k]- P_j[k-1]$, and $\Delta Q_j[k]\coloneqq Q_j[k]- Q_j[k-1]$. In the following, we initially treat the GFL inverters as uncontrollable resources, and neglect the changes in the total active and reactive power losses that result from the secondary control execution, assuming that $\sum_{j\in \mathcal{V}^{(I)}}\Delta P_j[k]\approx0$, and $\sum_{j\in \mathcal{V}^{(I)}}\Delta Q_j[k]\approx0$. Then, it follows from \eqref{voc_eq_lin1} that the weighted average of the inverter voltage magnitudes, denoted by $\overline{E}[k] =\sum_{j\in\mathcal{V}^{(I)}} D_{j,v}E_j[k]/\sum_{l\in\mathcal{V}^{(I)}} D_{l,v}$, changes as follows:
\begin{align}\label{average_voltage}
\Delta \overline{E}[k] &= \frac{1}{\sum_{j\in\mathcal{V}^{(I)}} D_{j,v}}\sum_{j\in\mathcal{V}^{(I)}}\Delta Q_j^r[k],
\end{align}
where $\Delta\overline{E}[k] \coloneqq \overline{E}[k]-\overline{E}[k-1]$.
By using \eqref{average_voltage}, and summing~\eqref{voc_eq_lin2} over all $j\in\mathcal{V}^{(I)}$ and rearranging, we obtain
\begin{subequations}\label{voc_eq_lin3}
\begin{align}
\Delta \overline{E}[k] &= R_{eq}^{(v)}\sum_{j\in \mathcal{V}^{(I)}}\Delta Q_j^r[k],\\
\Delta\omega[k]&= R_{eq}^{(f)}\sum_{j\in \mathcal{V}^{(I)}}\Delta P_j^r[k],
\end{align}
\end{subequations}
where $R_{eq}^{(v)} = 1/\sum_{j\in\mathcal{V}^{(I)}} 1/R_j^{(v)}$ and $R_{eq}^{(f)} = 1/\sum_{j\in\mathcal{V}^{(I)}} 1/R_j^{(f)}$. The constant $R_{eq}^{(f)}$ is commonly referred to as the composite frequency droop representing the combined effect of the frequency droops of all GFM-inverters. Similarly, we refer to $R_{eq}^{(v)}$ as the composite voltage droop. 
The relations \eqref{voc_eq_lin3} motivate the following initialization strategy for the sensitivity estimates:
\begin{subequations}\label{init}
\begin{align}
S_f &= [R_{eq}^{(f)}e^\top,\mathbf{0}_{m}^\top],
\mbox{ }S_v = \mathbf{1}_{d_v}[\mathbf{0}_{m}^\top,R_{eq}^{(v)}e^\top],\\
S_l &= \mathbf{0}_{d_l\times 2m},
\end{align}
\end{subequations}
where $e=[e_1,e_2,\dots,e_m]^\top$, with $e_i=1$, $i\in\mathcal{V}^{(I)}$, and $e_i=0$, otherwise, $\mathbf{0}_{m}$ denotes the all-zeros vector of length $m$, and $\mathbf{0}_{d_l\times 2m}$ denotes the $(d_l\times 2m)$-dimensional all-zeros matrix.
In the strategy \eqref{init}, GFM inverters are assumed to have the same impact on system frequency and voltage magnitudes, captured by the composite frequency and voltage droops, and have no effect on the line flows. Meanwhile, the sensitivity estimates corresponding to the GFL inverters are all set to zero implying that the PV arrays are initially treated as uncontrollable resources. Despite being inaccurate, the initialization strategy \eqref{init} provides us with an idea of how large the initial estimates should be.

\section{Data-Driven Secondary Control Design}\label{sec:sec_control}
\begin{figure}
    \centering
    \includegraphics[trim=0cm 0cm 0cm 0cm, clip=true, scale=0.7]{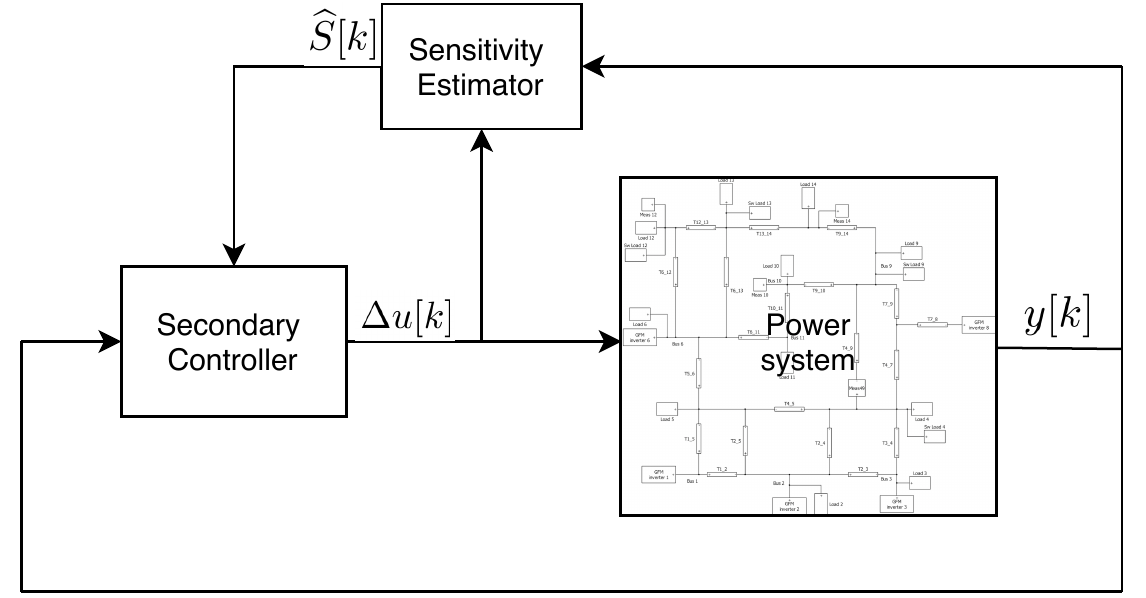}
    \caption{Diagram of the proposed data-driven secondary controller.}
    \label{fig:diagram}
    \vspace{-15pt}
\end{figure}

In this section, we present the secondary controller that relies on the output prediction model for regulating the system output to the nominal value, and the learned power-voltage characteristics of the PV arrays for dealing with uncertainty in solar power generation. Figure~\ref{fig:diagram} depicts the diagram of the proposed controller. The sensitivity estimator in \eqref{rls} is executed in parallel with the controller. Specifically, the applied control input adjustment and the measured resulting change in the output, $(\Delta u[k],\Delta y[k])$, are provided to the estimator to further update the sensitivity matrix estimate, $\widehat{S}[k]$, using \eqref{rls}. Controller performance depends on the accuracy of the prediction model decided by the richness of the control input adjustments. This motivates the strategy used in this paper to allow the proposed controller to generate persistently exciting control inputs.
\subsection{Feedback Optimization with Persistent Excitation}
In the following, we present the secondary controller as a solver of some underlying optimization problem.
We begin with the following optimization problem:
\begin{subequations}\label{sec_control0}
\begin{align}
\underset{\varphi}{\mbox{minimize}} & \mbox{ } \|\Delta y^*[k] - \widehat{S}[k]\varphi\|_2^2 + \rho\|\varphi\|_2^2\\
\mbox{subject to} & \mbox{ } \underline{\Delta u}[k] \leq \varphi \leq \overline{\Delta u}[k],\label{box_constr}
\end{align}
\end{subequations}
where $\rho\geq 0$, $\varphi=[\varphi_1,\varphi_2,\dots,\varphi_{2m}]^\top$, $\underline{\Delta u}[k]=[\underline{\Delta u}_1[k],\underline{\Delta u}_2[k],\dots,\underline{\Delta u}_{2m}[k]]^\top$, $\overline{\Delta u}[k]=[\overline{\Delta u}_1[k],\overline{\Delta u}_2[k],\dots,\overline{\Delta u}_{2m}[k]]^\top$, with $\underline{\Delta u}_i[k]$ and $\overline{\Delta u}_i[k]$ denoting respectively the lower and upper bounds on the active power setpoint adjustment at inverter $i\in\mathcal{V}^{(I)}$ at time instant $t_k$, and $\underline{\Delta u}_{i+m}[k]$ and $\overline{\Delta u}_{i+m}[k]$ denoting the lower and upper bounds on the reactive power setpoint adjustment at inverter $i\in\mathcal{V}^{(I)}$ at time instant $t_k$. The objective is to produce the output change $\widehat{S}[k]\varphi$ that tracks the target $\Delta y^*[k]\coloneqq y^*-y[k-1]$ while penalizing the control effort. The inequality constraints \eqref{box_constr} are meant for keeping the input adjustments within the incremental capacities of the inverters and inside the linear regime where the linear prediction model is accurate.
We then excite the control input adjustments by adding a random signal, denoted by $n[k]$, to the solution of the optimization problem \eqref{sec_control0}, denoted by $\varphi^*[k]$, as follows:
\begin{align}
\Delta u[k] = \varphi^*[k] + n[k].\label{sec_control_noise}
\end{align}
Despite the simplicity of the excitation approach, there exist several challenges in choosing suitable values for $n[k]$.
On the one hand, $n[k]$ needs to be large enough in order for its effect to be measurable at the output. On the other hand, if $n[k]$ is too large, the control input adjustments might push the operating point away from the desired one degrading the controller performance. Even more importantly, \eqref{sec_control_noise} does not take into account random variations in $\Delta y^*[k]$, e.g., due to load fluctuations, which can add a significant amount of excitation to $\varphi^*[k]$, making the excitation via the addition of $n[k]$ unnecessary. In what follows, we describe an alternative approach for exciting the secondary control commands, where we make use of the random variations in $\Delta y^*[k]$ as appropriate, and avoid unnecessary excitation that \eqref{sec_control_noise} might generate.

Consider the following modification of the optimization problem \eqref{sec_control0}, where we add a random signal, denoted by $w[k]$, to the target $\Delta y^*[k]$:
\begin{align*}
\underset{\varphi}{\mbox{minimize}} & \mbox{ } \|\Delta y^*[k] + w[k] - \widehat{S}[k]\varphi\|_2^2 + \rho\|\varphi\|_2^2\\
\mbox{subject to} & \mbox{ } \underline{\Delta u}[k] \leq \varphi \leq \overline{\Delta u}[k],
\end{align*}
where $w[k] = 0$, if $\{\Delta y^*[l]\}_{l=1}^k$ is persistently exciting, and $w[k] \sim (-\Delta y^*[k])U(0,a_1)$, with $a_1\in(0,1)$, otherwise, where $U(a,b)$ denotes the continuous uniform distribution over the interval $(a,b)$.
For the time being, we neglect the inequality constraints in the following discussion. It holds trivially that if $\widehat{S}[k]$ is invertible for all $k\geq 1$, then, adding $w[l]$ to the target $\Delta y^*[l]$ will generate persistently exciting sequence $\{\varphi^*[l]\}_{l=1}^k$. However, if $\widehat{S}[k]$ has more columns than rows, then, adding $w[k]$ will provide excitation in the subspace that is orthogonal to the null space of $\widehat{S}[k]$. This opens up the possibility of the sequence $\{\varphi^*[l]\}_{l=1}^k$ not being persistently exciting. To rule out such possibility, we excite the control input adjustments as follows:
\begin{align}
\Delta u[k] = \varphi^*[k], 
\end{align}
if $\{\Delta u[l]\}_{l=1}^k$ is persistently exciting,
and 
\begin{align}
\Delta u[k] = \varphi^*[k] + \alpha[k] \nu[k],
\end{align}
otherwise, where $\nu[k]\in\mathds{R}^{2m}$ and $\alpha[k]\in\mathds{R}$ are randomly chosen such that $\nu[k]\in\mbox{Null}(\widehat{S}[k])$\footnote{$\mbox{Null}(A)$ denotes the null space of an $m\times n$ matrix $A$, i.e., $\mbox{Null}(A) \coloneqq \{x:x\in\mathds{R}^n\mbox{ and }Ax = 0\}$} and
\begin{align}
\underline{\Delta u}[k] \leq \varphi^*[k]+ \alpha[k] \nu[k] \leq \overline{\Delta u}[k].\label{alpha}
\end{align} 
Since $\widehat{S}[k]\nu[k] = 0$, we note that the perturbation $\alpha[k] \nu[k]$ does not affect the predicted output change, $\widehat{S}[k]\varphi^*[k]$. Hence, if the prediction model is accurate, we expect the perturbation not to have any effect on the system output.

Next, we discuss the control excitation taking into account the inequality constraints. The inequality constraints might become active if the target $\Delta y^*[k]$ is sufficiently large. As a result, the control input adjustments may not be excitable without violating the inequality constraints, namely, there might not exist non-zero $\alpha[k]$ such that \eqref{alpha} holds.
To guard against such possibility, we keep the control input adjustments persistently exciting by randomly varying the lower and upper bounds of the constraints in \eqref{box_constr}, yielding the following optimization problem:
\begin{subequations}\label{sec_control}
\begin{align}
\underset{\varphi}{\mbox{minimize}} & \mbox{ } \|\Delta y^*[k] + w[k] - \widehat{S}[k]\varphi\|_2^2+ \rho\|\varphi\|_2^2\\
\mbox{subject to} & \mbox{ } \underline{\Delta u}[k] + \eta_1[k] \leq \varphi \leq \overline{\Delta u}[k] - \eta_2[k],
\end{align}
\end{subequations}
where $\eta_1[k] \sim U(0,a_2|\underline{\Delta u}[k]|)$, $\eta_2[k]\sim U(0,a_2|\overline{\Delta u}[k]|)$, with $a_2\in(0,1)$, at time instant $t_k$. Note that if all inequality constraints are active at the solution, then, the approach is equivalent to \eqref{sec_control_noise}. On the other hand, if only a subset of the inequality constraints are active, then, the target $\Delta y^*[k]$ plays a non-negligible role in exciting the solution $\varphi^*[k]$. The pseudocode for the proposed secondary control approach is provided in Algorithm~1.
\begin{figure}
\noindent\vsepfbox{%
\begin{varwidth}{\dimexpr\linewidth-2\fboxsep-2\fboxrule\relax}
\textit{Algorithm~1. Data-Driven Secondary Control with Persistent Excitation.}\\
1. Initialize $\rho\geq 0$, $a_1\in(0,1)$, $a_2\in(0,1)$.\\
2. Set $w[k] = 0$, if $\{\Delta y^*[l]\}_{l=1}^k$ is persistently exciting, and $w[k] \sim (-\Delta y^*[k])U(0,a_1)$, otherwise; $\eta_1[k] \sim U(0,a_2|\underline{\Delta u}[k]|)$; $\eta_2[k] \sim U(0,a_2|\overline{\Delta u}[k]|)$.\\
3. Find a solution of \eqref{sec_control}, $\varphi^*[k]$, and
apply
\begin{equation*}
\Delta u[k] = \varphi^*[k], 
\end{equation*}
if $\{\Delta u[l]\}_{l=1}^k$ is persistently exciting,
and 
\begin{equation*}
\Delta u[k] = \varphi^*[k] + \alpha[k] \nu[k],
\end{equation*}
otherwise, where $\nu[k]$ and $\alpha[k]$ are randomly chosen such that $\nu[k]\in\mbox{Null}(\widehat{S}[k])$ and
\begin{equation*}
\underline{\Delta u}[k] \leq \varphi^*[k]+ \alpha[k] \nu[k] \leq \overline{\Delta u}[k].
\end{equation*} 
\end{varwidth}%
}\vspace{-10pt}
\end{figure}
\subsection{Secondary Control with PV Power Tracking}
In the following, we take into consideration the constraints that PV arrays impose on the secondary controller during its execution. As discussed earlier, to extract a desired amount of power from a PV array, a certain voltage must be applied across its terminals, as determined by its power-voltage characteristic. Hence, the voltage setpoint must be computed so as to allow the PV array to track the secondary control inputs. 
To this end, we augment the optimization problem \eqref{sec_control} with the learned power-voltage characteristics \eqref{poly_model} as follows:
\begin{subequations}\label{sec_control_pv_tracking}
\begin{align}
\underset{\varphi,v}{\mbox{minimize}} & \mbox{ } \|\Delta y^*[k] +w[k]- \widehat{S}[k]\varphi\|_2^2+ \rho\|\varphi\|_2^2\\
\mbox{subject to} & \mbox{ } \underline{\Delta u}[k] + \eta_1[k] \leq \varphi \leq \overline{\Delta u}[k] - \eta_2[k],\\
& \mbox{ }\underline{\Delta u}[k] \leq \varphi+ \alpha[k] \nu[k] \leq \overline{\Delta u}[k],\\
& \mbox{ } c_i(v_i,\beta^{(i)}[k]) = P_i[k-1]+\varphi_i+\alpha_i[k] \nu_i[k],\label{power_volt_constr}\\
& \mbox{ }\underline{v}_{i} \leq v_{i} \leq \overline{v}_{i},\mbox{ } i \in \mathcal{V}_{pv},
\end{align}
\end{subequations}
where $v=[v_1,v_2,\dots,v_{|\mathcal{V}_{pv}|}]^\top$, $c_i(v_i,\beta^{(i)}[k])$ represents the estimated power-voltage characteristic of the PV array~$i$, with $\beta^{(i)}[k] = \big[\beta_0^{(i)}[k],\beta_2^{(i)}[k],\dots,\beta_{2d}^{(i)}[k]\big]^\top$ denoting the estimated coefficients of the fitted polynomial model, $\underline{v}_{i}$ and $\overline{v}_{i}$ denote the lower and upper bounds on the voltage applied across the terminals of the PV array~$i\in \mathcal{V}_{pv}$, and $P_i[k-1]\coloneqq P_i(t_{k}^-)$, with $P_i(t)$ denoting the active power injection by the PV array~$i\in \mathcal{V}_{pv}$ at time $t$. The constraint \eqref{power_volt_constr} serves to maintain a power balance between the generated solar power and the inverter active power output, while neglecting the inverter losses. Note that we have integrated step~4 of Algorithm~1 directly into the optimization \eqref{sec_control_pv_tracking}, where $\alpha[k]=0$, if $\{\Delta u[l]\}_{l=1}^{k-1}$ is persistently exciting, and $\alpha[k]$ is randomly chosen, otherwise. Due to the non-linearity of the power-voltage characteristics in \eqref{power_volt_constr}, the problem \eqref{sec_control_pv_tracking} is non-convex. However, as described next, it has an equivalent convex reformulation comprising two stages.

In the first stage, we consider the following convex relaxation of \eqref{sec_control_pv_tracking}, where the power-voltage characteristics are replaced with the inequality constraints that are meant to keep the control input adjustments within the incremental capacities of the PV arrays:
\begin{subequations}\label{sec_control_pv_tracking_convex}
\begin{align}
\underset{\varphi}{\mbox{minimize}} & \mbox{ } \|\Delta y^*[k] +w[k]- \widehat{S}[k]\varphi\|^2+ \rho\|\varphi\|_2^2\\
\mbox{subject to} & \mbox{ } \underline{\Delta u}[k] + \eta_1[k] \leq \varphi \leq \overline{\Delta u}[k] - \eta_2[k],\\
& \mbox{ }\underline{\Delta u}[k] \leq \varphi+ \alpha[k] \nu[k] \leq \overline{\Delta u}[k],\\
 & \mbox{ } \underline{\Delta u}^{pv}_i[k] \leq \varphi_i+\alpha_i[k] \nu_i[k] \leq \overline{\Delta u}^{pv}_i[k], \mbox{ } i \in \mathcal{V}_{pv},\label{pv_constr}
\end{align}
\end{subequations}
where $\overline{\Delta u}^{pv}_i[k] = \overline{p}_{i}[k]-P_i[k-1]$, and $\underline{\Delta u}^{pv}_i[k] = -P_i[k-1]$ denote the maximum and minimum incremental capacities of the PV array~$i$, and $\overline{p}_{i}[k] = \max\limits_{\underline{v}_{i} \leq v_i \leq \overline{v}_{i}}c_i(v_i,\beta^{(i)}[k])$ denotes the estimate of the active power capacity of the PV array~$i$. In \eqref{sec_control_pv_tracking_convex}, we are content with finding the power setpoints that are to be tracked by the PV arrays given their limited incremental capacities. To find the corresponding voltage setpoints, we execute the second stage as follows. We compute the voltage setpoint of the PV array $i \in \mathcal{V}_{pv}$, denoted by $v_i^*[k]$, such that $c_i(v_i^*[k],\beta^{(i)}[k]) = P_i[k-1]+\varphi_i^*[k]+\alpha_i[k]\nu_i[k]$, with $\varphi^*[k]$ denoting the solution of \eqref{sec_control_pv_tracking_convex}, by using the power tracking algorithm in \cite[Algorithm~2]{ZhDh20}, which is based on Newton's method.

Since it is easy to see that $\big(\varphi^*[k], v^*[k]\coloneqq\big[v_1^*[k],v_2^*[k],\dots,v_{|\mathcal{V}_{pv}|}^*[k]\big]^\top\big)$ is the solution of \eqref{sec_control_pv_tracking}, we omit the proof. Inclusion of the inequality constraints in \eqref{pv_constr} may lead to a lack of persistent excitation of the control inputs. Similar to \eqref{sec_control}, to guard against such possibility, we randomly vary the lower and upper bounds of the constraints in \eqref{pv_constr} if the incremental capacities of the PV arrays are not persistently exciting, yielding the following optimization problem:
\begin{subequations}\label{sec_control_pv_tracking_convex2}
\begin{align}
\underset{\varphi}{\mbox{minimize}} & \mbox{ } \|\Delta y^*[k] +w[k]- \widehat{S}[k]\varphi\|^2+ \rho\|\varphi\|_2^2\\
\mbox{subject to} & \mbox{ } \underline{\Delta u}[k] + \eta_1[k] \leq \varphi \leq \overline{\Delta u}[k] - \eta_2[k],\\
& \mbox{ }\underline{\Delta u}[k] \leq \varphi+ \alpha[k] \nu[k] \leq \overline{\Delta u}[k],\\
 & \mbox{ } \underline{\Delta u}^{pv}_i[k] + \underline{\zeta}_i[k] \leq \varphi_i+\alpha_i[k]\nu_i[k],\mbox{ }i \in \mathcal{V}_{pv},\nonumber\\&\mbox{ } \varphi_i+\alpha_i[k]\nu_i[k]\leq \overline{\Delta u}^{pv}_i[k] - \overline{\zeta}_i[k],
\end{align}
\end{subequations}
where $\underline{\zeta}_i[k]=0$, if $\{\underline{\Delta u}^{pv}[l]\}_{l=1}^k$ is persistently exciting, and $\underline{\zeta}_i[k]\sim U(0,a_3|\underline{\Delta u}^{pv}_i[k]|)$, otherwise, with $a_3\in(0,1)$, at time instant $t_k$; $\overline{\zeta}_i[k]=0$, if $\{\overline{\Delta u}^{pv}[l]\}_{l=1}^k$ is persistently exciting, and $\overline{\zeta}_i[k]\sim U(0,a_3|\overline{\Delta u}^{pv}_i[k]|)$, otherwise.
In Algorithm~2, we provide the pseudocode for the proposed approach combining the secondary control objectives with the PV power tracking objectives.
\begin{figure}
\noindent\vsepfbox{%
\begin{varwidth}{\dimexpr\linewidth-2\fboxsep-2\fboxrule\relax}
\textit{Algorithm~2. Data-Driven Secondary Control with PV Power Tracking.}
\begin{itemize}
\item[1.]Initialize $\rho\geq0$, $a_1\in(0,1)$, $a_2\in(0,1)$, $a_3\in(0,1)$.
\item[2.]Set
\begin{itemize}
\item[$\vcenter{\hbox{\tiny$\bullet$}}$] $w[k] = 0$, if $\{\Delta y^*[l]\}_{l=1}^k$ is persistently exciting, and $w[k] \sim (-\Delta y^*[k])U(0,a_1)$, otherwise;
\item[$\vcenter{\hbox{\tiny$\bullet$}}$] $\eta_1[k] \sim U(0,a_2|\underline{\Delta u}[k]|)$; $\eta_2[k] \sim U(0,a_2|\overline{\Delta u}[k]|)$;
\item[$\vcenter{\hbox{\tiny$\bullet$}}$] $\underline{\zeta}_i[k]=0$, if $\{\underline{\Delta u}^{pv}[l]\}_{l=1}^k$ is persistently exciting, and $\underline{\zeta}_i[k]\sim U(0,a_3|\underline{\Delta u}^{pv}_i[k]|)$, otherwise;
\item[$\vcenter{\hbox{\tiny$\bullet$}}$] $\overline{\zeta}_i[k]=0$, if $\{\overline{\Delta u}^{pv}[l]\}_{l=1}^k$ is persistently exciting, and $\overline{\zeta}_i[k]\sim U(0,a_3|\overline{\Delta u}^{pv}_i[k]|)$, otherwise;
\item[$\vcenter{\hbox{\tiny$\bullet$}}$] $\alpha[k]=0$, if $\{\Delta u[l]\}_{l=1}^{k-1}$ is persistently exciting, and $\alpha[k]$ is randomly chosen, otherwise;
\item[$\vcenter{\hbox{\tiny$\bullet$}}$] $\nu[k]$ is randomly chosen such that $\nu[k]\in\mbox{Null}(\widehat{S}[k])$.
\end{itemize}
\item[3.] Find a solution of \eqref{sec_control_pv_tracking_convex2}, $\varphi^*[k]$,
and apply $\Delta u[k] = \varphi^*[k]$.
\item[4.] Compute $v_i^*[k]$ such that $c_i(v_i^*[k],\beta^{(i)}[k]) = P_i[k-1]+\varphi_i^*[k]+\alpha_i[k]\nu_i[k]$, $i \in \mathcal{V}_{pv}$, by using \cite[Algorithm~2]{ZhDh20}.
\end{itemize}
\end{varwidth}%
}\vspace{-12pt}
\end{figure}
\section{Numerical Simulations}
\begin{figure}
    \centering
    \includegraphics[width=1\columnwidth]{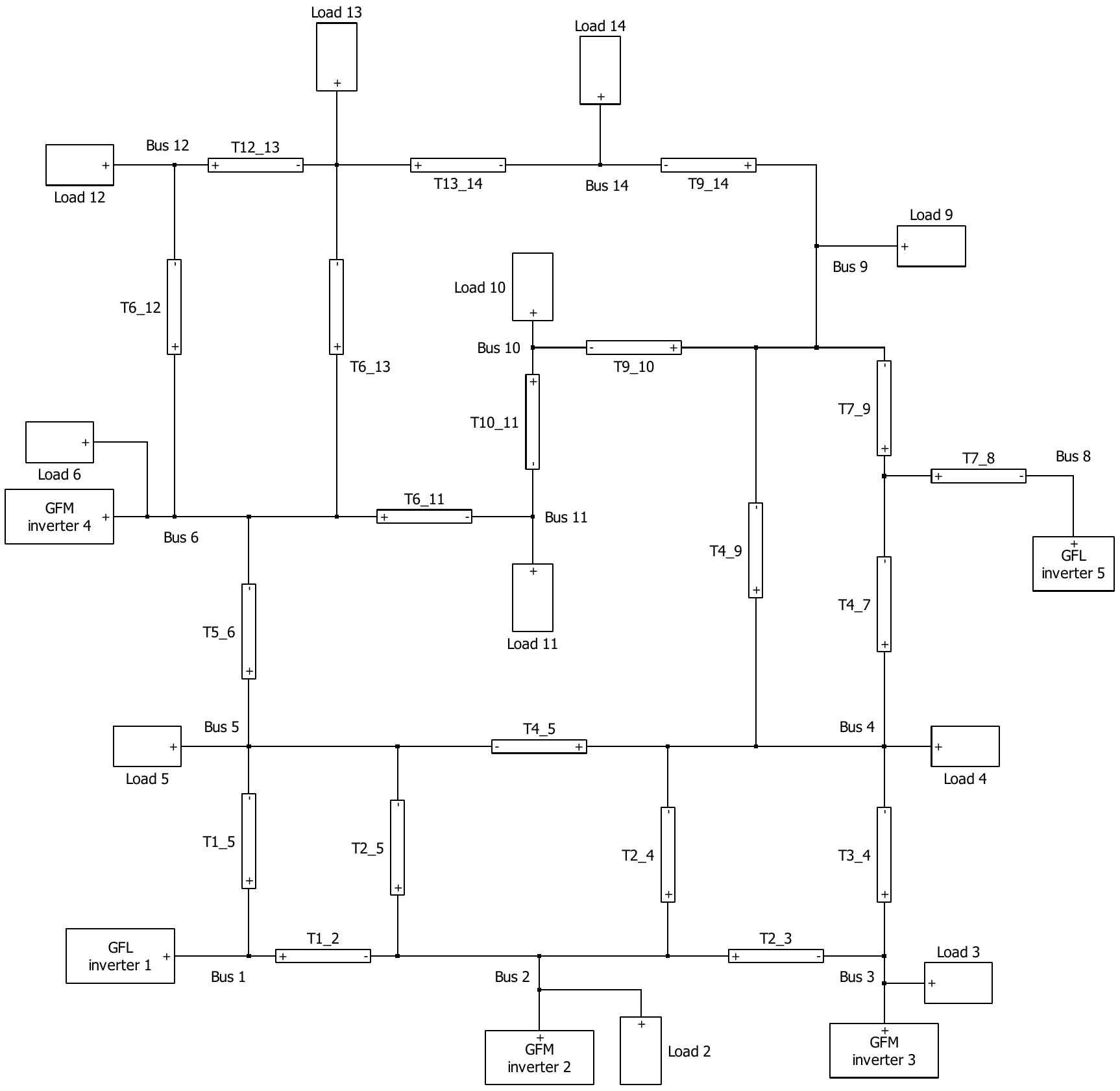}
    \caption{The single line diagram of the modified IEEE-14 bus test system.}
    \label{fig:IEEE14}
    \vspace{-20pt}
\end{figure}

\begin{figure}
    \centering
    \begin{tabular}{c}
    \includegraphics[trim=0cm 0cm 0cm 0cm, clip=true, scale=0.3]{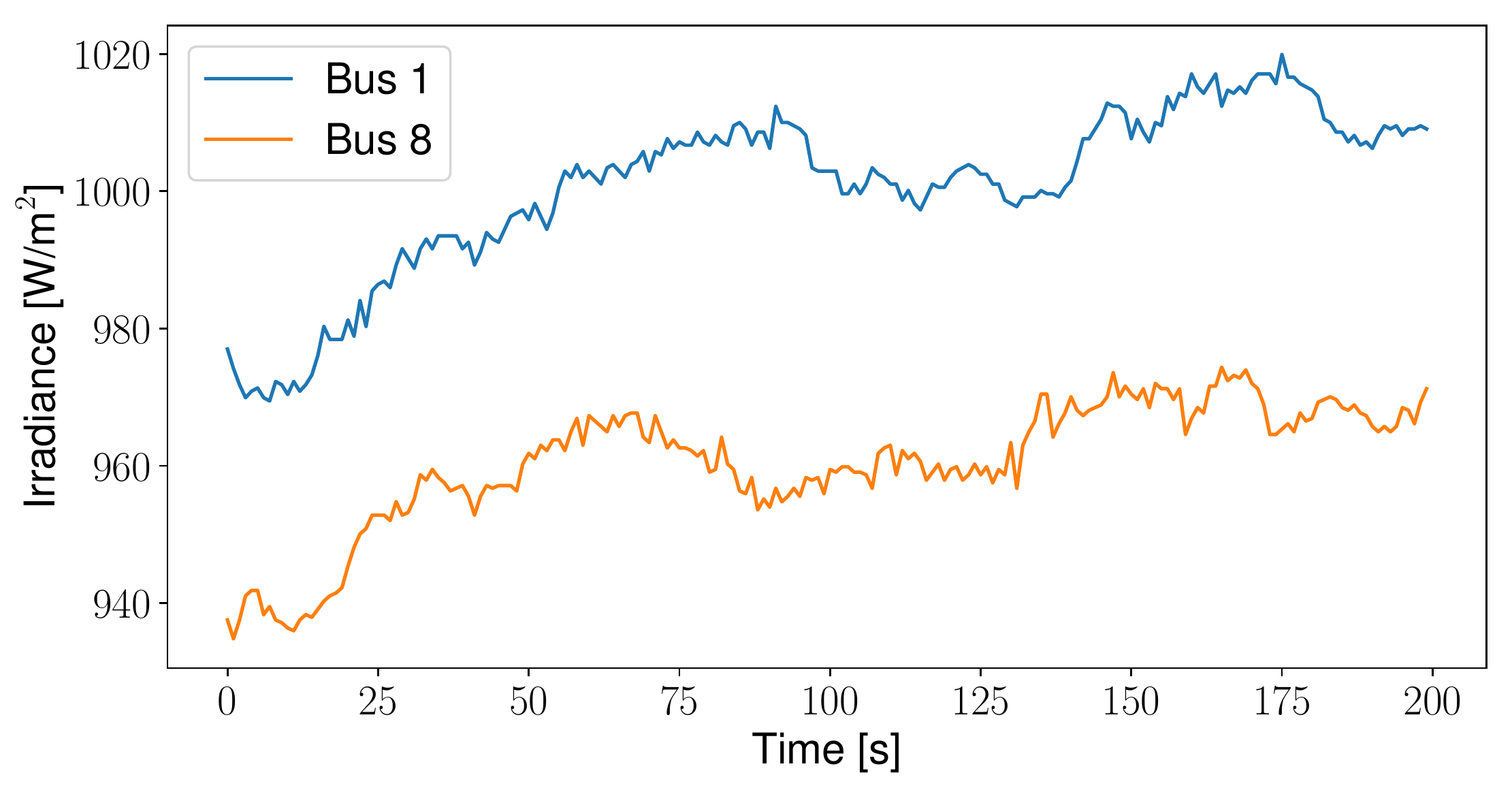}\\
    \includegraphics[trim=0cm 0cm 0cm 0cm, clip=true, scale=0.3]{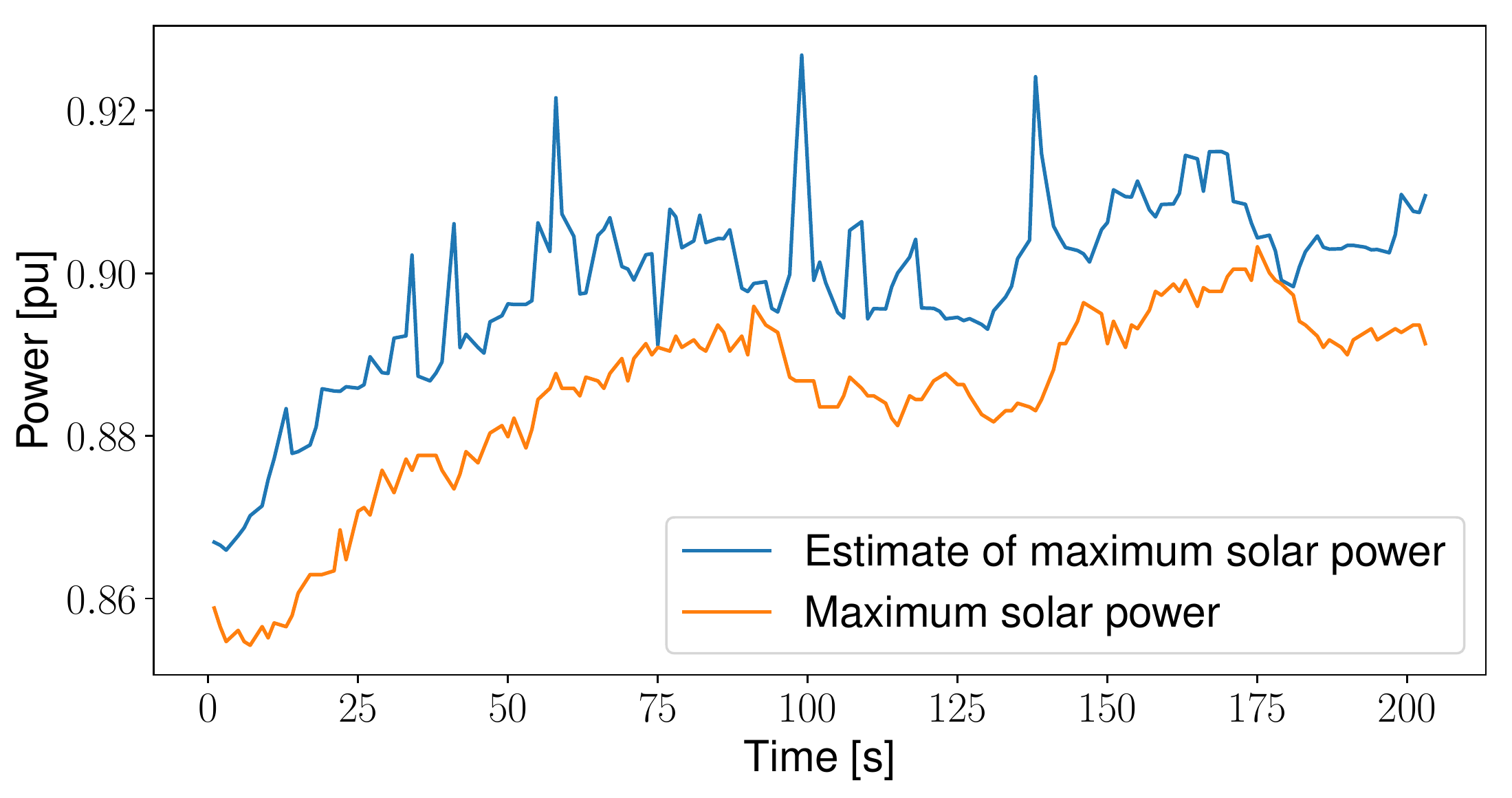}
    \end{tabular}
    \caption{Solar irradiance profile at buses $1$ and $8$, maximum available solar power at bus~$1$ and its estimate.}
    \label{fig:solar}
    \vspace{-15pt}
\end{figure}

We showcase the proposed secondary controller using a modified version of the IEEE-14 bus test system \cite{IEEE14_test}, with energy sources providing both active and reactive power at buses $1$, $2$, $3$, and reactive power energy sources at buses $6$ and $8$ replaced by inverter-interfaced DERs, as described in the single line diagram of the test system in Figure~\ref{fig:IEEE14}. The GFM-inverter-interfaced DERs are located at buses $2$, $3$ and $6$, and the GFL inverters interfacing PV arrays are located at buses $1$ and $8$. The DER index sets are given by: $\mathcal{V}_{pv}=\{1,5\}$, and $\mathcal{V}^{(I)}=\{2,3,4\}$.
PV generation is simulated at buses $1$ and $8$ using solar irradiance data collected from NREL's Oahu Solar Measurement Grid \cite{Oahu_NREL_solar_data} on Oct 1, 2011, from 11:56:40~am to 12:00:00~pm; see Figure~\ref{fig:solar}. As described in Section~\ref{subsec:pv_tracking}, the estimation of the power-voltage characteristics of the PV arrays at buses $1$ and $8$ is performed by fitting the polynomial models in \eqref{poly_model}, with $d=4$, to the collected power-voltage measurements $\{(v_1[l],p_1[l])\}_{l=k-w}^k$ and $\{(v_5[l],p_5[l])\}_{l=k-w}^k$, with $w=9$. Figure~\ref{fig:solar} also shows the maximum available solar power at bus~$1$ and its estimate obtained from the learned power-voltage characteristics.

Load perturbations occur every~$4$~s, and the secondary controller is executed every~$1$~s except for the time instants when the loads are perturbed. This separation between the controller actions and load changes is introduced to clearly demonstrate the effectiveness of the control approach. Otherwise, if the loads change while the system is settling down after a secondary controller execution, it is more challenging to evaluate the controller performance and the accuracy of the prediction model.

We execute Algorithm~1 to determine the control input adjustments, and use the following values to initialize its parameters: $\rho=0$, $a_1 = 0.5$, $a_2 = 0.4$, $\underline{\Delta u}_i[k]=-0.1$~pu, $\overline{\Delta u}_i[k]=0.1$~pu, $i\in\mathcal{V}^{(g)}$, and $\alpha[k]\sim U(0,0.05)$. The feedback optimization problem \eqref{sec_control_pv_tracking_convex} is solved using CVXPY \cite{cvxpy}. The sensitivity estimator in \eqref{rls} is executed in parallel with the controller. Specifically, the applied control input adjustment and the measured resulting change in the output, $(\Delta u[k],\Delta y[k])$, are provided to the estimator to further update the sensitivity matrix estimate, $\widehat{S}[k]$, using \eqref{rls}. The estimator is initialized using \eqref{init}. We also use the following values to initialize the estimator parameters in \eqref{rls}: $\rho_1=1000$, $\lambda=0.85$.

\begin{figure}
    \centering
    \begin{tabular}{r}
    \includegraphics[trim=0cm 0cm 0cm 0cm, clip=true, scale=0.3]{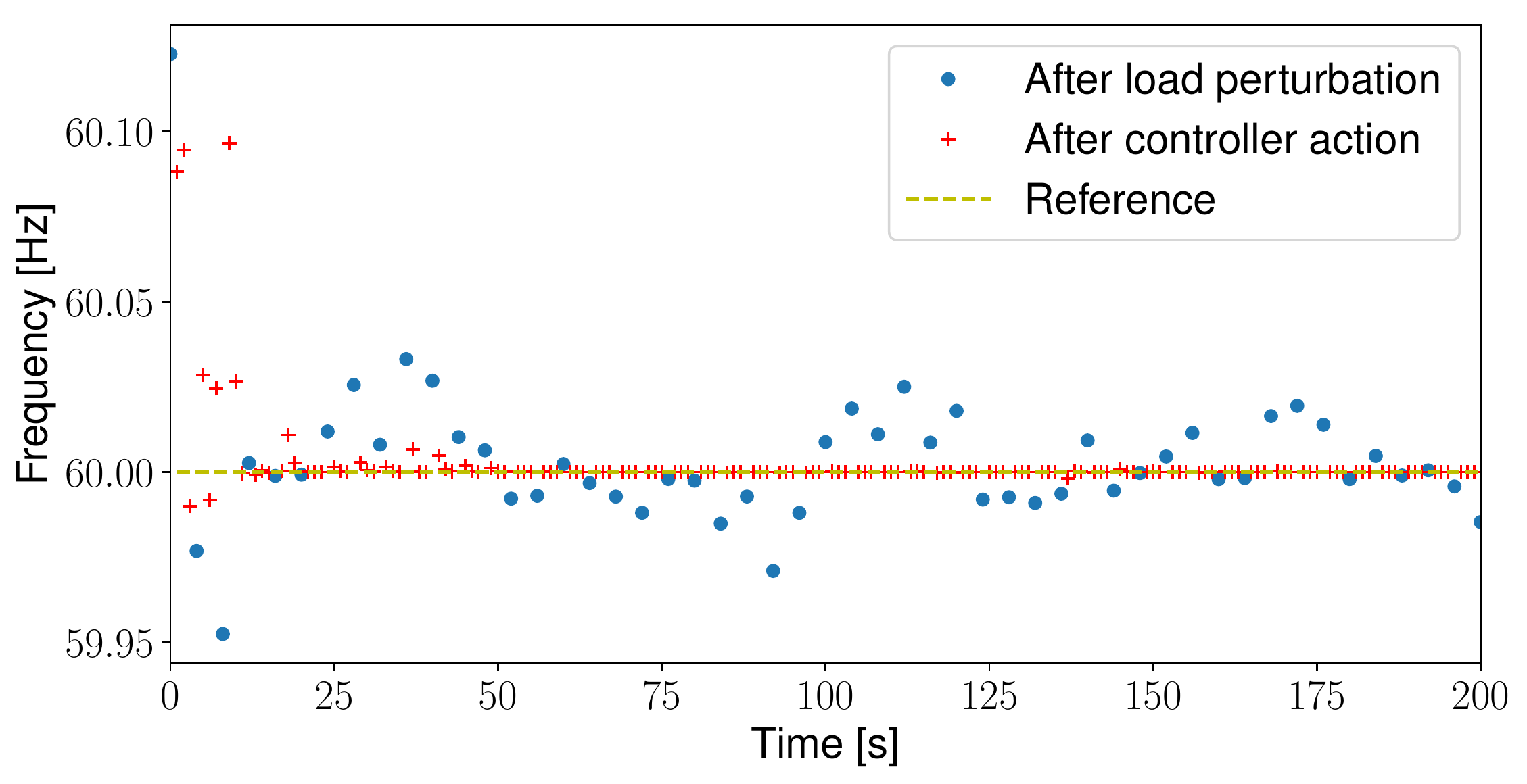}\\[-5pt]
    \includegraphics[trim=0cm 0cm 0cm 0cm, clip=true, scale=0.3]{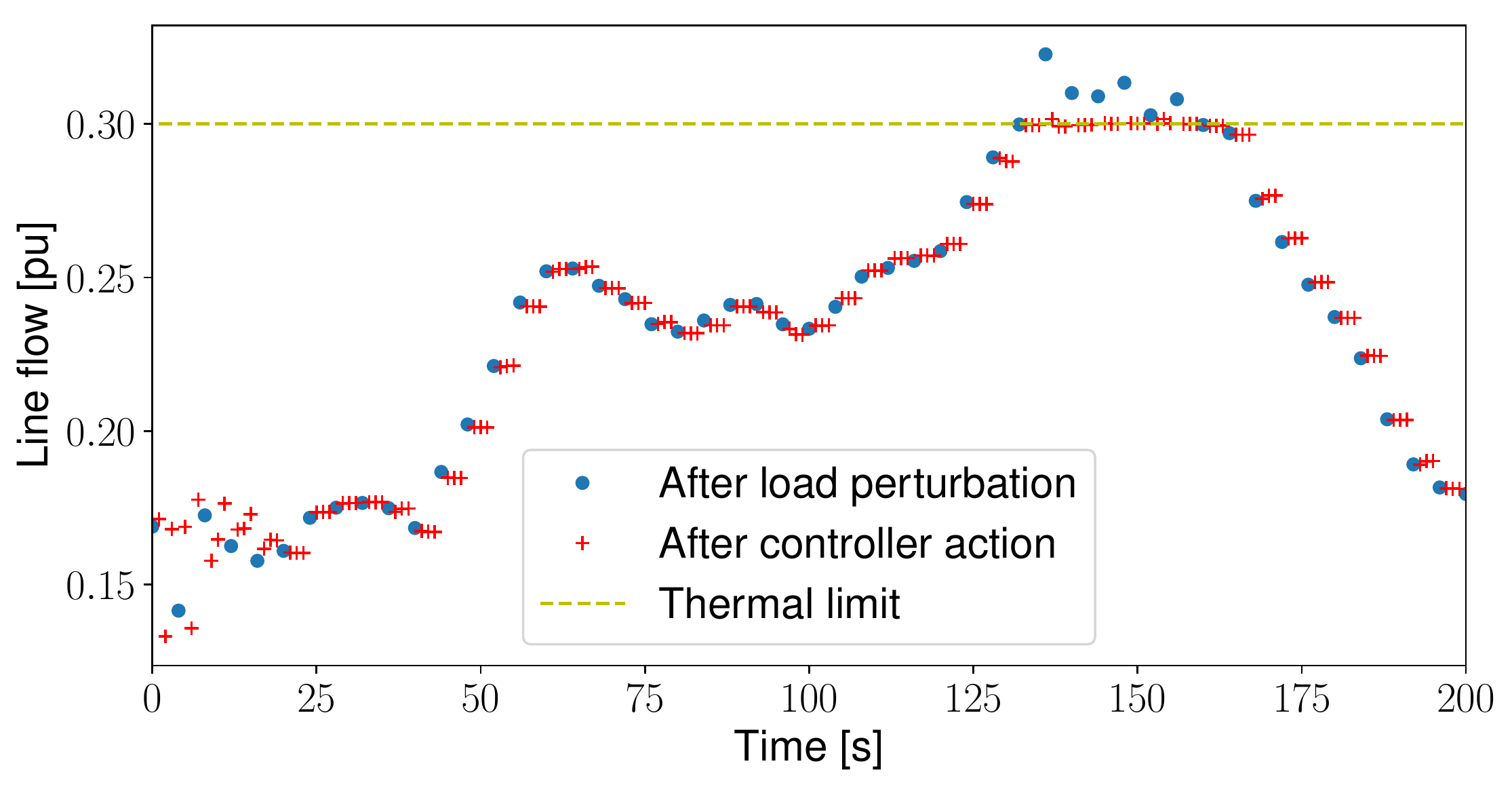}\\[-5pt]
    \includegraphics[trim=0cm 0cm 0cm 0cm, clip=true, scale=0.3]{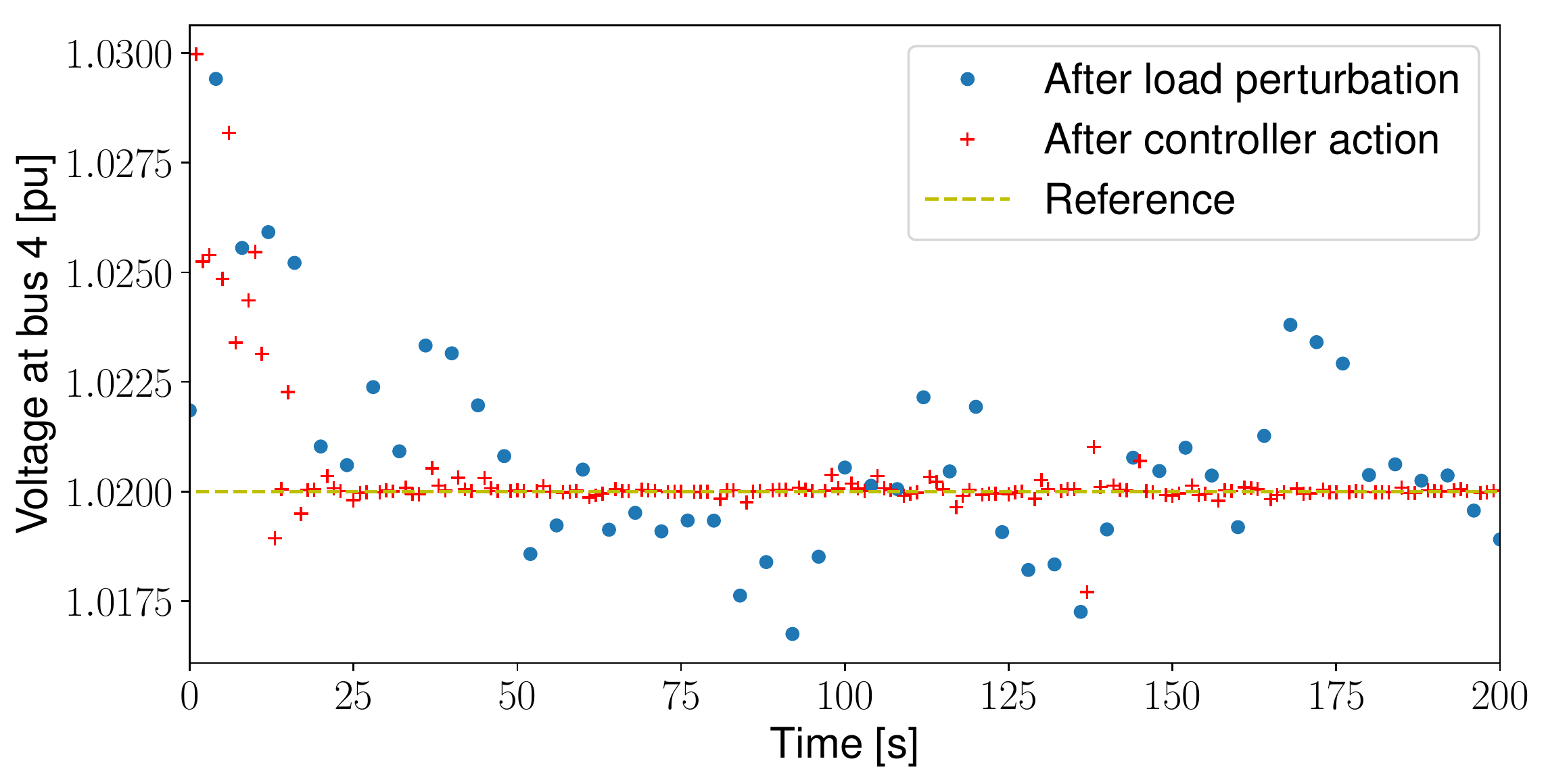}\\[-5pt]
    \includegraphics[trim=0cm 0cm 0cm 0cm, clip=true, scale=0.3]{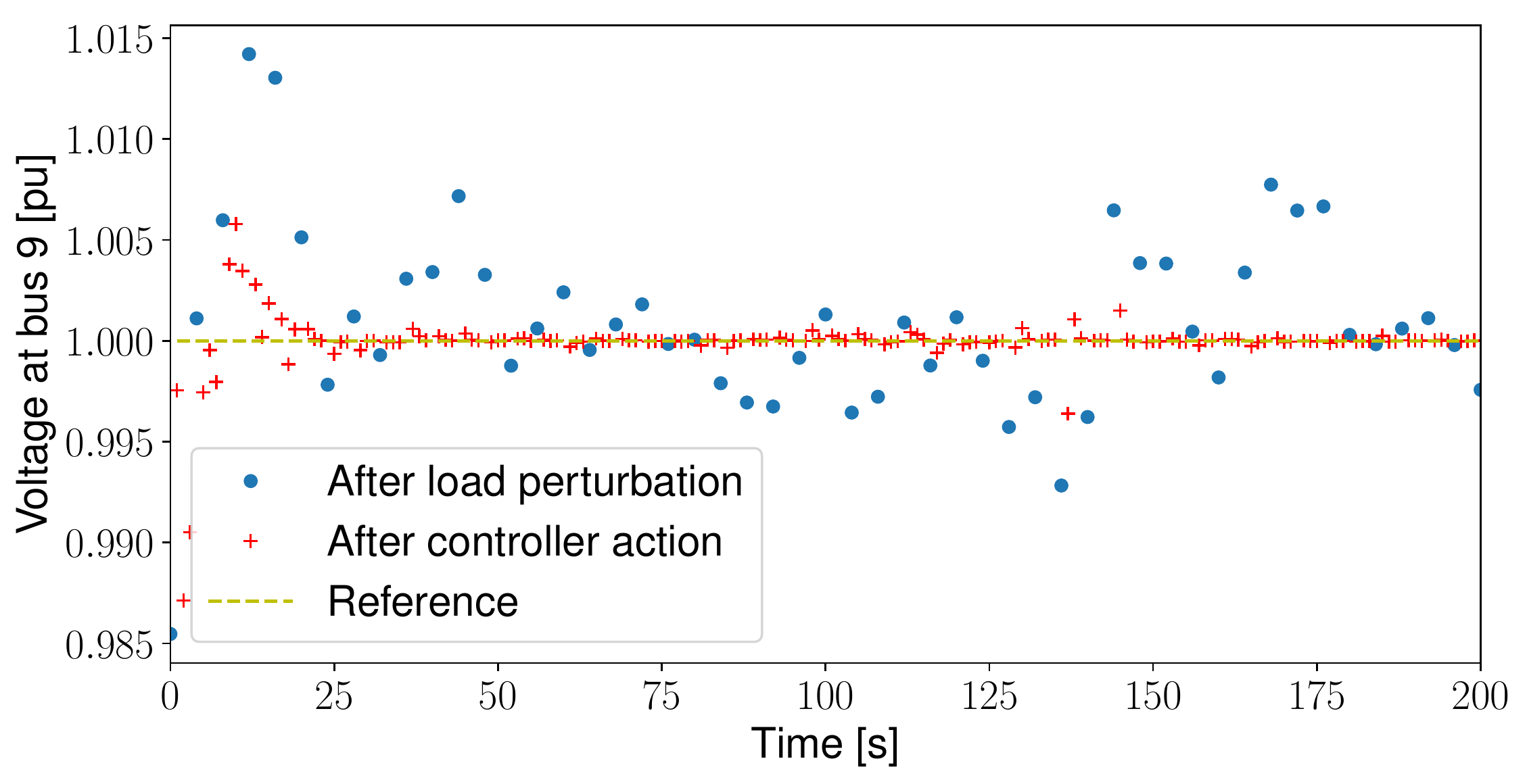}\\[-5pt]
    \includegraphics[trim=0cm 0cm 0cm 0cm, clip=true, scale=0.3]{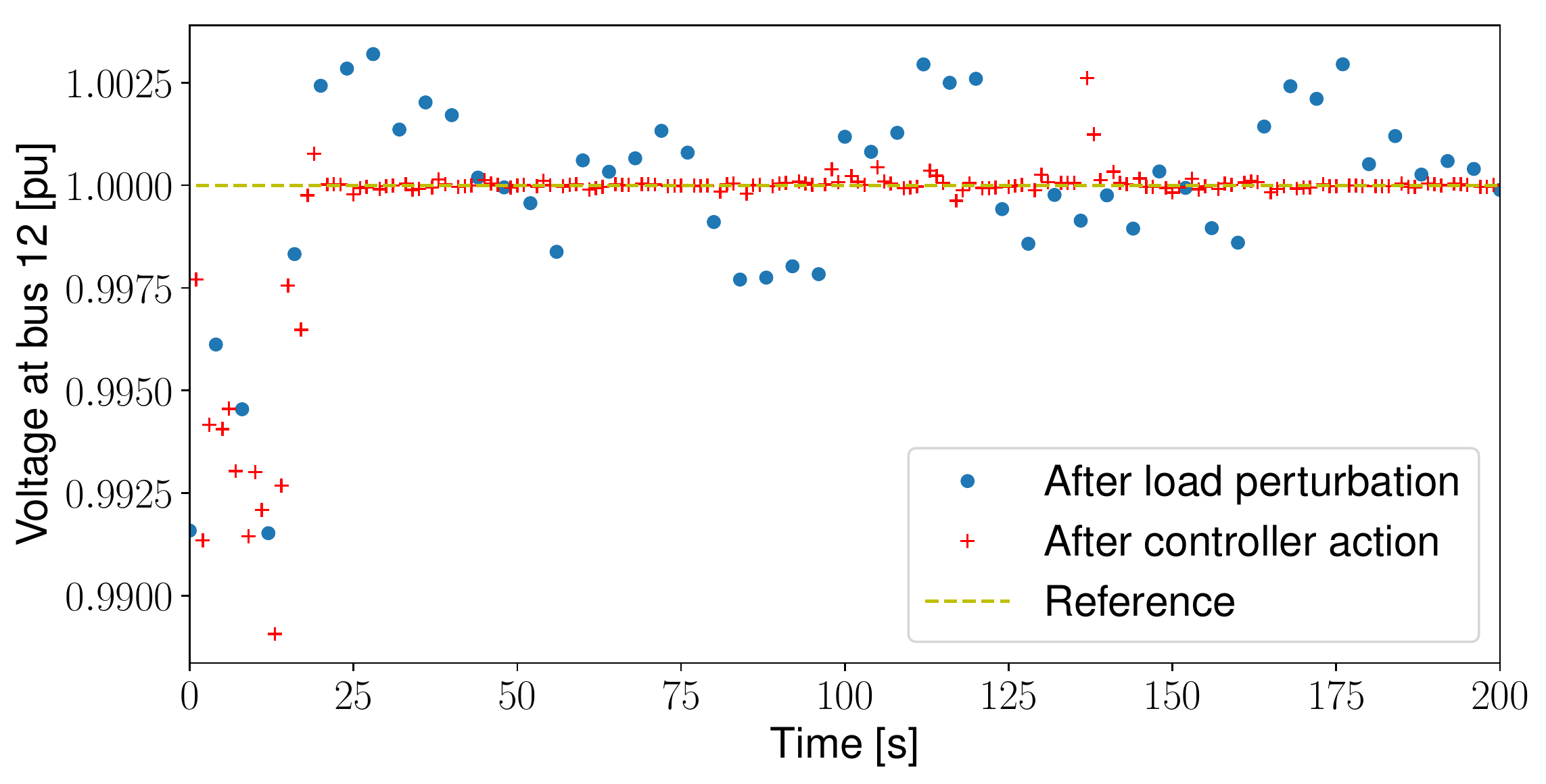}
    \end{tabular}
    \caption{Trajectories of the system frequency, active power flow through line $4-9$, voltage magnitudes at buses $4$, $9$, and $12$.}
    \label{fig:freq_volt}
    \vspace{-15pt}
\end{figure}

Figure~\ref{fig:freq_volt} shows the evolution of the system output, namely, the system frequency and voltage magnitudes at buses $4$, $9$, $12$ being regulated to the corresponding nominal values, $60$~Hz, $V_{4}^*=1.02$~pu, $V_{9}^*=1$~pu, and $V_{12}^*=1$~pu, and the active power flow through line $4-9$ kept below its thermal limit set at $0.3$~pu. To be more specific, this figure shows the steady-state values of the system output after each load perturbation, and after each secondary control action. [Note that three control actions are taken before another load perturbation occurs.] It can be seen that the control actions determined by Algorithm~1 manage to keep the system output close to the nominal values despite the system experiencing significant load variations. We note that the control actions do not always bring the system output noticeably closer to their nominal values, because the controller constantly makes a tradeoff between various objectives---frequency regulation, line flow regulation, and voltage regulation---while satisfying the constraints on the control input adjustments.

\begin{figure}
    \centering
    \begin{tabular}{r}
    \includegraphics[trim=0cm 0cm 0cm 0cm, clip=true, scale=0.3]{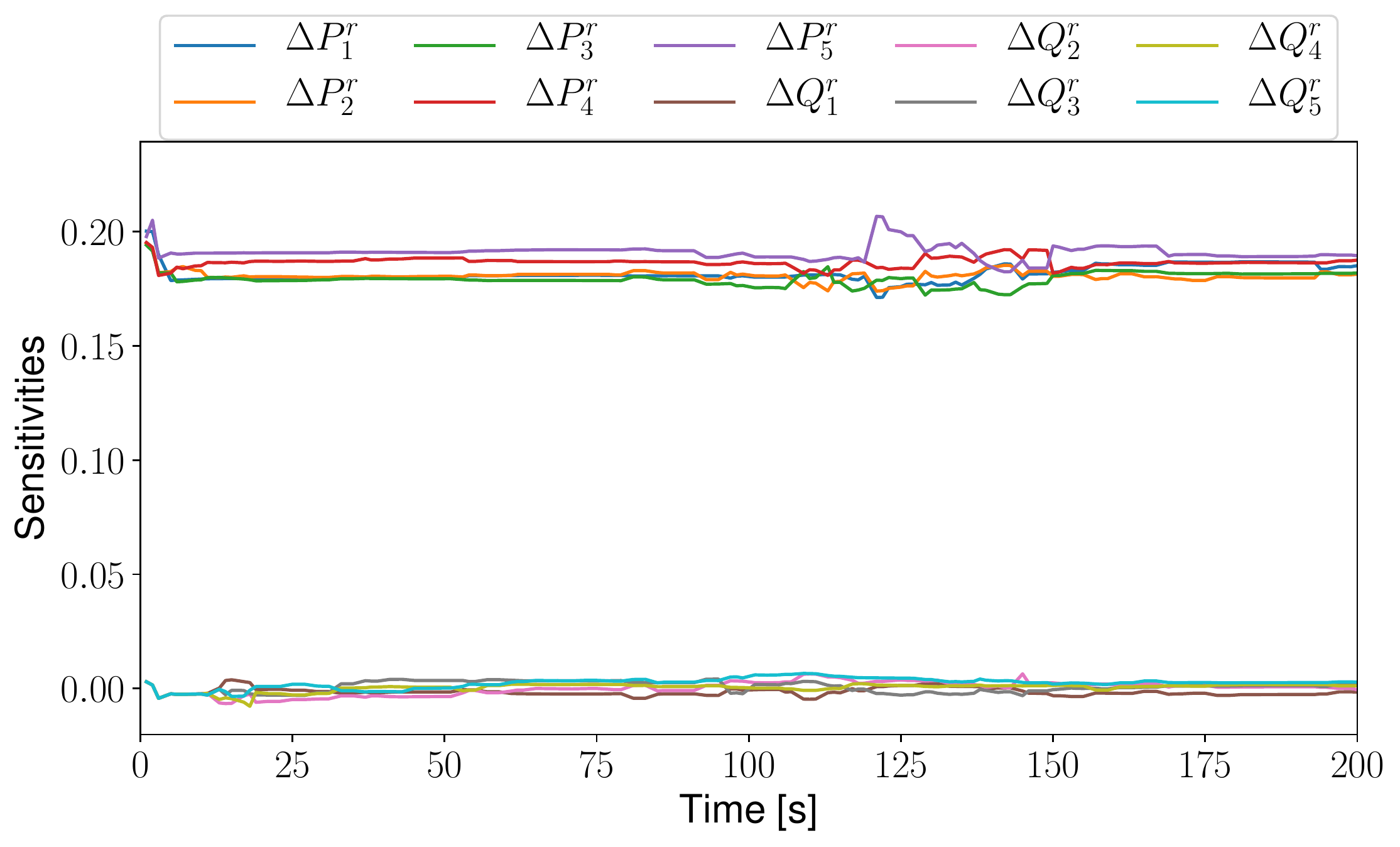}\\
    \includegraphics[trim=0cm 0cm 0cm 0cm, clip=true, scale=0.3]{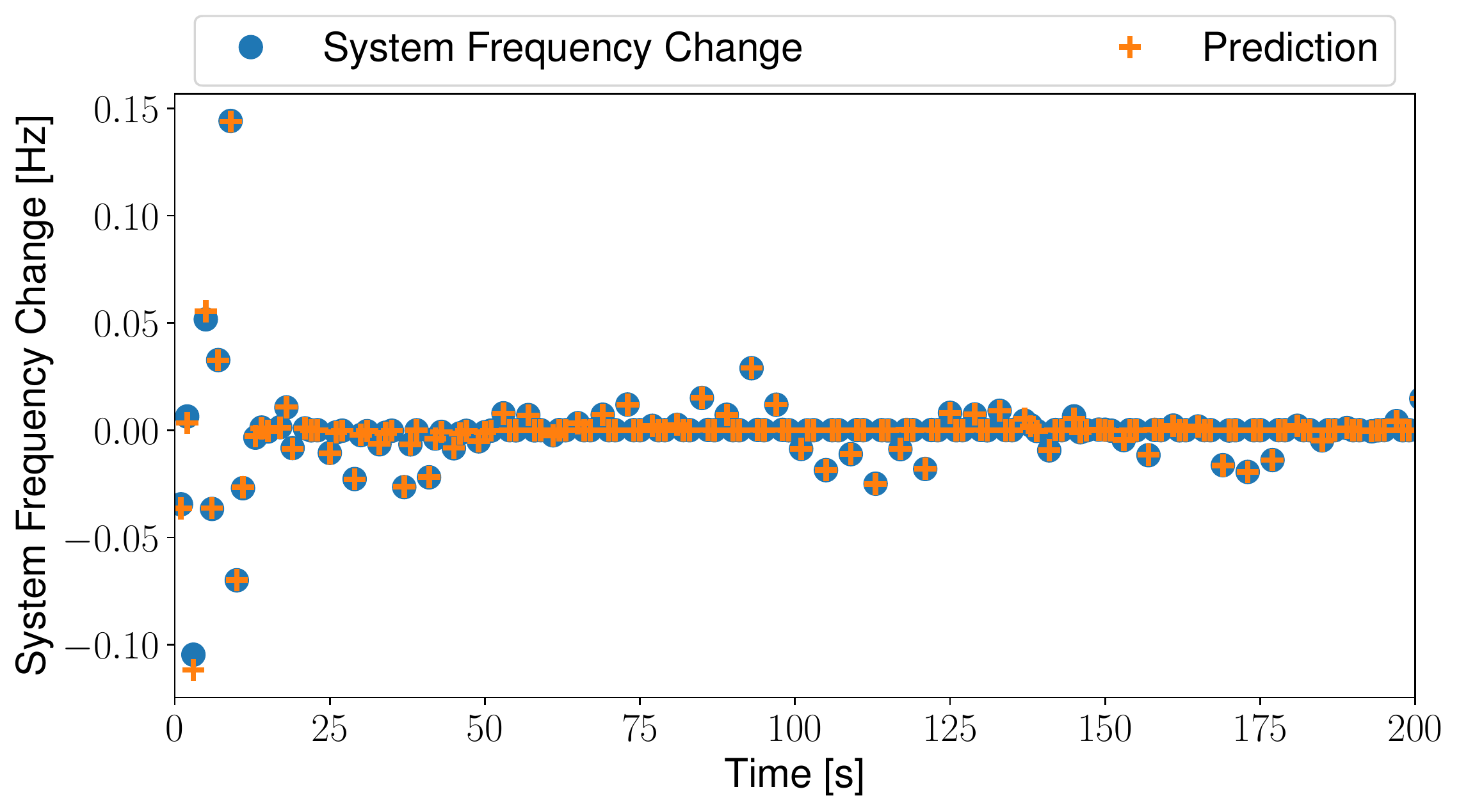}
    \end{tabular}
    \caption{Evolution of the sensitivities corresponding to the system frequency, and predictions of the frequency changes.}
    \label{fig:prediction_results}
    \vspace{-15pt}
\end{figure}

Figure~\ref{fig:prediction_results} shows the evolution of the estimates of the sensitivities corresponding to the system frequency. During the first $20$~s, the sensitivity estimates experience some transients quickly reaching a steady state. Note that the initial estimates are not far away from the steady-state ones, which explains the satisfactory performance of the controller in regulating frequency during the transient period. Also note that the active power setpoint adjustments, $\Delta P^r[k]$, have much more significant effect on the frequency than the reactive power setpoint adjustments, $\Delta Q^r[k]$, which is consistent with the inductive nature of the test system network.
The smallest and largest eigenvalues of $\sum_{l=k-\tau}^{k} \Delta u[l]\Delta u[l]^\top$, with $\tau=20$, are bounded and strictly positive and equal, on average, $0.0025$ and $0.54$, demonstrating that $\{\Delta u[l]\}_{l=1}^k$ satisfies \eqref{persistent_excitation}, and, thus, is persistently exciting.
Figure~\ref{fig:prediction_results} also shows the predictions of the system frequency made by the linear prediction model based on the learned sensitivities. Prediction errors are less than $10$~\% in $80-90$~\% of all test cases, where the prediction errors are computed as follows:
\[e_i[k] = \frac{|\Delta y_i[k]-\widehat{\Delta y}_i[k]|}{\max(|\Delta y_i[k]|,\overline{y}_{m,i})}100\%,\]
where $e_i[k]$ is the prediction error corresponding to the output change $\Delta y_i[k]$, $\overline{y}_{m,i}$ is the threshold below which the changes in the output $y_i$ are not measurable. For example, for the system frequency, we set $\overline{y}_{m,1} = 10^{-3}$~Hz. We note that the predictions are less accurate during the initial transient period, when the operating point changes significantly, or when the control input adjustments are too large going over the linear regime of the prediction model.

\section{Concluding Remarks and Future Work}
In this paper, we developed a data-driven secondary controller for power networks with a deep penetration of GFM inverters. The controller, which is based on feedback optimization, relies on learning specific power system sensitivities, e.g., of the system frequency to the inverter setpoints, for predicting the system output changes given the input adjustments. It is critical to keep the feedback optimization solution persistently exciting in order to be able to make accurate predictions. To this end, we presented an approach that maintains persistent excitation of the control inputs without significantly degrading the controller performance. 

One of the future directions that we intend to explore is to utilize a more sophisticated prediction model in the feedback optimization to improve prediction accuracy. However, in the case of standard polynomial models or linear transfer function models (see, e.g., \cite{AW:08}), predictions were less accurate in our simulations.

\bibliographystyle{IEEEtran}
\bibliography{References}
\end{document}